\documentclass[conference]{IEEEtran}
\IEEEoverridecommandlockouts

\usepackage{cite}
\usepackage{amsmath,amssymb,amsfonts}
\usepackage{algorithmic}
\usepackage{graphicx}
\usepackage{textcomp}
\usepackage{xcolor}
\usepackage{comment}
\usepackage{colortbl}
\usepackage{booktabs}

\usepackage{xurl}
\makeatletter
  \long\def\@makecaption#1#2{%
    \ifx\@captype\@IEEEtablestring
      \footnotesize\bgroup\par\centering\@IEEEtabletopskipstrut
        {\normalfont\footnotesize #1:\nobreakspace #2}
      \par\egroup%
      \@IEEEtablecaptionsepspace
    \else
      \@IEEEfigurecaptionsepspace
      \setbox\@tempboxa\hbox{\normalfont\footnotesize {#1.}\nobreakspace\nobreakspace #2}%
      \ifdim \wd\@tempboxa >\hsize
        \setbox\@tempboxa\hbox{\normalfont\footnotesize {#1.}\nobreakspace\nobreakspace}%
        \parbox[t]{\hsize}{\normalfont\footnotesize\noindent\unhbox\@tempboxa#2}%
      \else
        \ifCLASSOPTIONconference
          \hbox to\hsize{\normalfont\footnotesize\hfil\box\@tempboxa\hfil}%
        \else
          \hbox to\hsize{\normalfont\footnotesize\box\@tempboxa\hfil}%
        \fi
      \fi
    \fi}
  \makeatother

\usepackage{titlesec}
\titleformat{\paragraph}[runin]
{\normalfont\normalsize\bfseries}{}{0pt}{}
\titlespacing*{\paragraph}{\parindent}{0ex}{1em}

\def\BibTeX{{\rm B\kern-.05em{\sc i\kern-.025em b}\kern-.08em
    T\kern-.1667em\lower.7ex\hbox{E}\kern-.125emX}}
\begin{document}

\title{WikiKV: Schema-Evolving Path-Indexed Storage for Hierarchical Knowledge Navigation\\

}
\author{
  \textbf{Feifei Li \textsuperscript{*}},
  \textbf{Haoliang Ming \textsuperscript{*}},
  \textbf{Zihan Li \textsuperscript{*}},
  \textbf{Hang Liao},\\
  \textbf{Xingyu Fan},
  \textbf{Xiaoqing Wu},
  \textbf{Chenggong Wang},
  \textbf{Wenhui Que}\textsuperscript{\dag}
\\
  WeChat, Tencent Inc., Beijing, China
\\
  \small\texttt{\{niyali, hliangming, muselli, caryliao\}@tencent.com}\\
  \small\texttt{\{fanxfan, xiaoqingwwu, jevanwang, victorque\}@tencent.com}
\\[2pt]
  \textsuperscript{*}These authors contributed equally to this work.\\
  \textsuperscript{\dag}Corresponding author.
}

\maketitle

\begin{abstract}

LLM-curated hierarchical knowledge bases, namely a tree-structured wiki whose nodes summarize an underlying corpus, have become a dominant substrate for retrieval-augmented applications, yet their storage layer is still treated as an implementation detail. This workload is hierarchical, query-intensive, and continuously evolving, and no existing storage model natively captures all three properties at once.

We present \textsc{WikiKV}, a path-indexed key--value storage model purpose-built for this workload, comprising three components: (i) a data-driven schema that bootstraps the hierarchy via Intent-Anchored Schema Induction and refines it through Continuous Evolution Operators; (ii) a consistency protocol for the path-indexed storage model that precludes partial-read observations under concurrent offline rewrites without read-path locking; and (iii) a budgeted navigation operator whose search-accelerated routing reduces the expected number of LLM-assisted descent steps from $d$ to $O(1)$ while preserving anytime semantics with progressively refined answers.
We evaluate \textsc{WikiKV} through real-world deployment for the WeChat Official Account AI Assistant and benchmark it against diverse baselines on the \textsc{AuthTrace} dataset, where it achieves balanced low per-operator latency across four query operators against relational, graph, and FS backends, and reaches $63.2\%$ end-to-end answer correctness, exceeding multiple RAG baselines, with the gap widening on low- and high-fan-in multi-document questions. Ablation study further confirms the effectiveness of \textsc{WikiKV}'s components.

\end{abstract}


\section{Introduction}
\label{sec:introduction}

At the intersection of large language models (LLMs)~\cite{openai2023gpt4,yang2024qwen2,grattafiori2024llama3} and data systems, a new paradigm of knowledge management is emerging. Rather than indexing flat document collections for retrieval-augmented generation (RAG)~\cite{lewis2020rag,gao2023ragsurvey}, production applications increasingly compile their unstructured corpora into \emph{hierarchical knowledge bases}, tree-structured wiki whose nodes summarize and organize the underlying material---echoing the curated structures of community-built knowledge bases such as Wikidata~\cite{vrandecic2014wikidata}, DBpedia~\cite{auer2007dbpedia}, and Freebase~\cite{bollacker2008freebase}. End-user queries are then answered by \emph{navigating} the hierarchy from a global index through dimensions down to individual entity pages and documents, often through agentic reasoning--acting loops over the structure~\cite{yao2023react}. This pattern already underlies LLM-curated assistants, enterprise knowledge portals, and content authoring tools, placing the storage layer on the critical path of every online query.

Three properties distinguish this workload from both document retrieval and general graph processing. First, access is \emph{hierarchical}: every read traverses the fixed schema \texttt{Index $\rightarrow$ Dimension $\rightarrow$ Entity $\rightarrow$ Digest $\rightarrow$ Document}. Second, the workload is \emph{read--write asymmetric}: online traffic is read-only under tight latency SLAs, while writes are batched into offline pipelines that construct and evolve the wiki. Third, the knowledge base is \emph{continuously evolving}: pages are added, merged and corrected as the corpus grows and access statistics expose stale content. A storage system for this paradigm thus needs low-latency path lookups and directory listings, snapshot-consistent reads under concurrent offline rewrites, and first-class schema evolution.

Yet no existing storage model natively supports this access pattern. Relational databases~\cite{stonebraker1986postgres} require recursive \texttt{JOIN}s or CTEs for deep paths, turning $O(1)$ navigation into $O(\textit{depth})$, and their fixed schemas cannot accommodate diverse wiki structures. Graph databases~\cite{robinson2018graphdatabases} like Neo4j model nodes and edges but lack a ``directory listing'' primitive, mismatching the read-page-and-enumerate-children contract with high operational overhead. Flat key-value stores~\cite{decandia2007dynamo,lakshman2010cassandra,chang2008bigtable} offer $O(1)$ lookups but no hierarchical semantics, forcing scans or secondary indexes for enumeration. Hierarchical file systems~\cite{ghemawat2003gfs,shvachko2010hdfs} fit structurally but are optimized for large sequential byte streams rather than fine-grained, queryable directory metadata under concurrent random reads, and they expose directory listings as opaque enumerations rather than as a first-class, payload-bounded query primitive. Practitioners therefore compose ad-hoc stacks of vector indexes~\cite{johnson2021faiss,wang2021milvus}, relational metadata, and bespoke caches, none of which captures the hierarchical semantics end to end.

We introduce \textsc{WikiKV}, a \emph{path-indexed} key--value (KV) storage model purpose-built for hierarchical knowledge bases. \textsc{WikiKV} encodes the wiki schema directly into the key space, so that one directory listing is served by a single point lookup on the directory node, in $O(1)$ storage round-trips. Above this storage core, \textsc{WikiKV} provides (i) a data-driven schema layer that initializes the hierarchy from corpus samples and continuously evolves it via merge and split operators, complemented by a content-level Error Book for cross-batch self-correction, (ii) a consistency protocol that prevents partial reads under write-while-read traffic, and (iii) a budgeted navigation query operator that uses search-accelerated routing to compress multi-step descents into $O(1)$ LLM-assisted hops while preserving progressive answer quality.

\textsc{WikiKV} presents the database-engineering realization of our broader LLM-curated wiki program, whose application-level retrieval method we introduced in our prior work~\cite{llmwiki}; that method achieved state-of-the-art results on multiple public benchmarks (see~\cite{llmwiki} for details). This paper focuses on the storage, deployment, and evolution side. Both the application-level retrieval behavior and the content-level Error Book are carried over from our prior work~\cite{llmwiki}, and are here re-grounded on the storage layer; the contributions of this paper are the database-layer components (path-as-key storage, the consistency protocol, the three-tier cache, and the schema-optimization formalization), and the end-to-end study of \S\ref{subsec:end-to-end} reports the full pipeline running on \textsc{WikiKV} to verify that the storage layer preserves its answer quality.

\textsc{WikiKV} has been successfully deployed in production to serve the WeChat Official Account AI Assistant, organizing WeChat Official Account articles into personal knowledge bases that enable the AI Assistant to answer followers' questions based on it.

Based on the calculation that each account contains hundreds to thousands of articles and knowledge pages, the system can support millions of account authors in building their personal knowledge bases.

Our main contributions are summarized as follows:
\begin{itemize}
  \item \textbf{Data-driven schema design and evolution.} We formalize hierarchical wiki construction as a constrained schema-optimization problem, and present algorithms for cold-start initialization and for continuous evolution via mutual-information-driven merge and Architect--Critic--Arbiter split, complemented by a content-level Error Book that persists across full and incremental ingestion runs.
  \item \textbf{Path-indexed storage model.} We propose a path-as-key encoding together with a structured value schema, and prove that under the accompanying parent-after-child write protocol every online query observes a consistent, partial-read-free view of the wiki. Both point lookups and directory listings are served in $O(1)$ storage round-trips.
  \item \textbf{Search-accelerated navigation queries.} We define a budgeted navigation query \textsc{NAV}$(q,B)$ with progressive semantics and show that search-accelerated routing reduces the expected number of LLM-driven navigation steps from $O(\textit{depth})$ to $O(1)$ for single-target queries, while still returning a usable coarse-grained answer whenever the budget is exhausted early.

\end{itemize}

We show that \textsc{WikiKV} outperforms representative baselines in per-operator latency across relational, graph, and FS backends, and achieves state-of-the-art end-to-end answer correctness against RAG baselines on the \textsc{AuthTrace} dataset.

The remainder of this paper is organized as follows.
Section~\ref{sec:preliminaries} formalizes the hierarchical knowledge-base model, the query model, and the consistency requirements. Section~\ref{sec:schema-evolution} describes data-driven schema cold-start and continuous evolution. Section~\ref{sec:storage-model} presents the \textsc{WikiKV} storage model and proves its consistency guarantees. Section~\ref{sec:navigation-query} introduces the budgeted navigation query and its search-accelerated execution plan. Section~\ref{sec:experiments} reports the experimental study. Section~\ref{sec:related-work} surveys related work, and Section~\ref{sec:conclusion} concludes.

\section{Preliminaries and Problem Formulation}
\label{sec:preliminaries}


\subsection{Hierarchical Knowledge Base Model}
\label{subsec:hkb-model}

A \emph{hierarchical knowledge base}, hereafter a \emph{wiki}, is an ordered rooted tree $T = (V, E)$ whose vertex set is partitioned into five ordered \emph{levels}, $V = V_I \cup V_D \cup V_E \cup V_{DI} \cup V_{DO}$, corresponding to the node types \emph{Index}, \emph{Dimension}, \emph{Entity}, \emph{Digest}, and \emph{Document}. The unique root $r \in V_I$ is the global index, internal nodes belong to $V_D$, and leaves belong to $V_E \cup V_{DI} \cup V_{DO}$. Every node $v \in V$ carries three pieces of state: a \emph{path} $\pi(v)$, a \emph{content} payload $c(v)$, and a \emph{metadata} record $m(v)$. The path $\pi(v) = /d_1/d_2/\cdots/d_n$ is the unique sequence of edge labels on the directed path from $r$ to $v$, encoded as a slash-separated string; it serves both as a human-readable address and, in §\ref{sec:storage-model}, as the storage key. We bound every node by $depth(v) \leq D$ for a depth budget $D$. The schema has five node types (Index, Dimension, Entity, Digest, Document); in our deployment each occupies one level, so $D$ = 5. $D$ may take any value, but effectiveness favors a small $D$ (typically 5). 
We take $D$ = 5 by default because the five node types already span the full \emph{index$\rightarrow$dimension$\rightarrow$entity$\rightarrow$digest$\rightarrow$document} abstraction chain, a larger $D$ only inserts extra Dimension routing levels that improve neither coverage nor answer quality
while lengthening every navigation descent. The storage model itself is depth-agnostic, but the operators in §\ref{sec:navigation-query} rely on this bound for their latency analysis.

\subsection{Query Model}
\label{subsec:query-model}

\begin{figure}[htbp]
    \centering
    \includegraphics[width=\columnwidth]{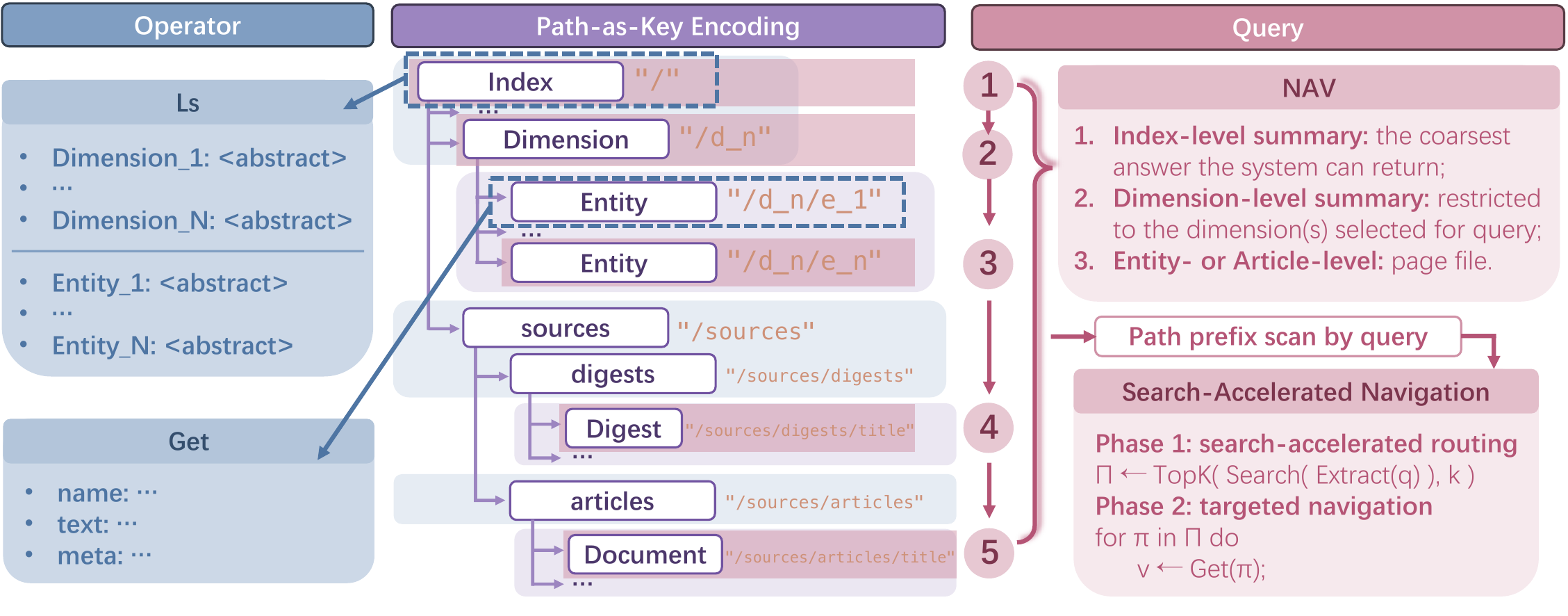}
    \caption{Four query operators over a hierarchical knowledge base (the physical KV key stored in the underlying engine is the hash digest $H(\pi(v))$).}
    \label{fig:query}
\end{figure}

Real workloads on hierarchical knowledge bases reduce to four operators that together capture the access patterns of LLM-curated knowledge applications as shown in Figure \ref{fig:query}:
\begin{itemize}
  \item \textbf{Q1: Path lookup.} $\textsc{Get}(\pi) \rightarrow v$ returns the node addressed by path $\pi$, or $\bot$ if no such node exists.
  \item \textbf{Q2: Directory list.} $\textsc{Ls}(\pi) \rightarrow (v, \langle \pi_1, \ldots, \pi_k \rangle)$ returns the node at $\pi$ together with the ordered list of paths of its children, where $k$ is the fan-out of $v$.
  \item \textbf{Q3: Navigation query.} $\textsc{Nav}(q, B) \rightarrow \langle r_1, r_2, \ldots, r_m \rangle$ takes a natural-language query $q$ and a time budget $B$, and returns a sequence of result records produced by descending the hierarchy under the budget.
  \item \textbf{Q4: Prefix search.} $\textsc{Search}(p) \rightarrow \{\pi : p \text{ is a prefix of } \pi\}$ returns the set of paths whose textual prefix matches $p$; it is used to accelerate the entry step of \textsc{Nav}.
\end{itemize}

The semantics of \textsc{Nav} differ from standard tree traversal in one essential way that we exploit in §\ref{sec:navigation-query}.

\paragraph*{Property 1 (Progressive answers)}
If $\textsc{Nav}(q, B)$ is interrupted after step $i$, the prefix $\langle r_1, \ldots, r_i \rangle$ remains a valid, albeit coarser-grained, answer to $q$. Equivalently, results are emitted in order of monotonically increasing granularity, so any prefix of the output is itself a usable answer.

\subsection{Consistency Requirements}
\label{subsec:consistency-requirements}

A wiki is read by online query traffic and written by offline construction-and-evolution pipelines. Any storage layer that realizes the above model must offer the following guarantees.

\textbf{R1: Read-after-write consistency.} Once a new page $v$ is admitted into the wiki, every subsequent directory listing $\textsc{Ls}(\pi(\mathrm{parent}(v)))$ returned to online queries must include $\pi(v)$, so that $v$ is reachable through navigation.

\textbf{R2: Concurrency safety under read--write asymmetry.} Online traffic is strictly read-only, and the construction pipeline is the sole writer. The storage layer must guarantee that readers never observe partial-write states, ideally without taking explicit locks; page-level incremental updates issued by the pipeline must respect the same property.

\textbf{R3: Bounded staleness.} After an offline write completes, online readers must observe its effect within a bounded staleness window $\Delta$, after which the new state is universally visible.

These three requirements decouple online read availability from the cost and complexity of offline writes. Section~\ref{sec:storage-model} gives a write protocol and a query-side fallback that together discharge R1--R3 without locking the read path.

\ifcsname c@algorithm\endcsname\else\newcounter{algorithm}\fi

\section{Data-Driven Schema Design and Evolution}
\label{sec:schema-evolution}

We address the dual challenges of cold-starting wiki schemas from corpora and enabling their continuous evolution. We formulate schema design as a constrained optimization problem, develop intent-anchored induction algorithms for initialization, and introduce two schema-evolution operators with a monotone-improvement guarantee, complemented by a content-level Error Book for cross-batch self-correction. These components integrate into an offline construction-and-evolution pipeline that supports dynamic schema adaptation.

\subsection{Motivation}
\label{subsec:schema-motivation}

Production knowledge bases outgrow hand-curation: even a well-designed five-level schema falls out of equilibrium once the corpus crosses an order of magnitude in size. Empirically, a wiki without a principled re-shaping discipline is unmaintainable past $10^4$ KV pairs in our deployments. We observe three concurrent forms of drift. \emph{Width drift} adds new top-level dimensions, forcing every navigation query to discriminate among many \textit{Index} siblings at the first hop and inflating both LLM routing difficulty and index-primitive constants. \emph{Depth-and-density drift} grows pages and edges roughly linearly with the corpus, bloating the materialized KV footprint and pushing directory listings past workable fan-out. \emph{Quality drift} accumulates stale content, low-confidence assertions, and never-read entries that raise the retrieval noise floor without contributing to recall. 


\subsection{Schema Design as a Constrained Optimization Problem}
\label{subsec:schema-formulation}

We treat schema design as a global optimization problem over the space of valid wikis of depth at most $D$ (five node types: Index, Dimension, Entity, Digest, Document; $D$ = 5 in our system, as in  \S\ref{subsec:hkb-model}). Let $S = (V, \ell, E)$ denote a candidate schema, where $V$ is its node set, $\ell : V \to \{I, D, E, \mathrm{DI}, \mathrm{DO}\}$ assigns each node to one of the five node types, and $E$ encodes the parent--child structure. Let $\rho$ be the access distribution that the online query workload induces over $V$, and let $W$ denote the workload itself in the form of \S\ref{subsec:query-model}. We score every candidate schema by the cost
\begin{equation}
\mathcal{C}(S; W) \;=\; \alpha\, |V| \;+\; \beta \sum_{v \in V} \mathrm{depth}(v)\cdot \rho(v) \;-\; \gamma\, \mathcal{Q}(S; W),
\label{eq:schema-cost}
\end{equation}
subject to a \emph{structural} constraint $\mathrm{depth}(v) \le D$ together with a per-node fan-out bound $|\mathrm{children}(v)| \le k_{\max}$.
Content-level quality is enforced separately by the Error Book repair loop introduced later in this section, rather than as a hard schema constraint.

Each term in (\ref{eq:schema-cost}) has a direct counterpart elsewhere in the paper. The storage term $\alpha\,|V|$ measures the size of the materialized KV namespace whose layout we will fix in Section~\ref{sec:storage-model}. The descent-depth term $\beta \sum_v \mathrm{depth}(v)\rho(v)$ is the access-weighted traversal cost that dominates online navigation latency on tightly budgeted queries. The quality term $\mathcal{Q}(S; W)$ is the end-to-end answer correctness that the workload exposes empirically (Section~\ref{sec:experiments}). The three coefficients $\alpha, \beta, \gamma > 0$ are deployment-time hyperparameters that trade storage footprint, online latency, and answer correctness against one another.

The full optimization is super-exponential in $|\mathcal{D}|$, and we do not attempt a global solution. Instead, we adopt a \emph{greedy local search} that starts from a cold-start schema $S_0$ (\S\ref{subsec:schema-cold-start}) and repeatedly applies the local operators of \S\ref{subsec:evolution-operators}, each of which we will show in Theorem~1 to monotonically decrease $\mathcal{C}$ under suitably chosen thresholds.

\subsection{Cold-Start: Intent-Anchored Schema Induction}
\label{subsec:schema-cold-start}

The cold-start problem is the zeroth-order instance of schema design: given a fresh corpus $\mathcal{D}$ and no prior structural assumptions, produce a valid initial schema $S_0$ that satisfies the structural and constraints of \S\ref{subsec:schema-formulation}. We solve it with a procedure we call \emph{Intent-Anchored Schema Induction} (IASI) that invokes an LLM directly and depends on \emph{no} statistical clustering, vector retrieval, or embedding model. The procedure runs once at deployment time, is decoupled from any online critical path, and consumes a sample $\mathcal{S} \subset \mathcal{D}$ whose size is fixed independently of $|\mathcal{D}|$.

\paragraph*{Intent-Anchored Schema Induction (IASI) procedure.}
Firstly, IASI consumes the sample $\mathcal{S}$ and emits a \emph{corpus positioning descriptor}
$$\mathcal{P} \;=\; \langle\, \mathrm{focus},\ \mathrm{audience},\ \mathrm{ingestion\mbox{-}bias}\,\rangle,$$
a short structured record stating what the corpus is about, who it is written for, and the bias under which documents enter the corpus. Then, IASI takes $(\mathcal{S}, \mathcal{P})$ as input and emits the directory scaffold $T$ that fixes $V_I$, $V_D$, $V_E$ and the parent--child structure of $E$ at these levels (digest and document nodes are populated dynamically by the ingestion pipeline); it carries the structural constraints of \S\ref{subsec:schema-formulation}, so that the tree is syntactically and structurally valid by construction on the first call rather than rejected and re-sampled by a post-hoc \emph{generate-then-validate} loop.


By contrast, the most direct cold-start baseline, which hands a sample of articles to an LLM with a prompt saying ``please produce a directory'' and parses the response, yields a directory anchored on whichever salient entities happen to surface in the first few sample documents, with long-tail entities absorbed into the fallback bucket. Rather than being a transient intermediate string consumed and discarded by the LLM, $\mathcal{P}$ is a first-class schema object materialized to durable storage alongside the directory tree, and read directly by subsequent evolution operators of \S\ref{subsec:evolution-operators}.

\paragraph*{Non-uniform sampling.}
The sample $\mathcal{S}$ is not drawn uniformly from $\mathcal{D}$. 
A lightweight \emph{ingestion filter} $\Phi$ runs before sampling and removes seven categories of low-information documents, such as boilerplate seasonal greetings, verbatim re-publications of upstream content, and event announcements.
Since $\mathcal{P}$ is constructed from LLM observations in $\mathcal{S}$, abundant low-information documents can systematically bias $\mathcal{P}$ toward inappropriate abstractions. Filtering before sampling prevents this miscalibration at the source instead of correcting it later.

\subsection{Continuous Evolution Operators}
\label{subsec:evolution-operators}

The cold-start schema $S_0$ drifts away from the optimum of \S\ref{subsec:schema-formulation} as soon as the corpus grows, the access distribution $\rho$ shifts, or page quality degrades. We define two local operators that move $S$ down the cost surface of $\mathcal{C}$ one step at a time, each consuming statistics that the storage layer already carries on every page (Section~\ref{sec:storage-model})---no external usage log or analytics warehouse is required.

\paragraph*{Operator~1: \textsc{DimensionMerge} (mutual-information-driven).}
For two sibling internal nodes $v_1, v_2$ that share a common parent, define the per-query co-access indicators $X_i = \mathbb{1}[\text{a query touches } v_i]$ and the mutual information
\begin{equation}
\mathrm{MI}(v_1, v_2) \;=\; \sum_{x_1, x_2} p_{12}(x_1, x_2)\, \log \frac{p_{12}(x_1, x_2)}{p_1(x_1)\, p_2(x_2)},
\label{eq:mi}
\end{equation}
estimated directly from the \texttt{access\_count} statistics co-located with each page record. When $\mathrm{MI}(v_1, v_2) > \theta_{\mathrm{merge}}$, we merge $v_1$ and $v_2$ into a single node $v_{12}$: the child list is the union of the originals' child lists, the \texttt{access\_count} is the sum, and the content is the concatenation of the originals' summaries. The operational interpretation is that two siblings whose access patterns are highly mutual are evidence that the user mental model treats them as a single concept; keeping them apart only widens the index and raises the routing burden of the navigation operator.

\paragraph*{Operator~2: \textsc{PageSplit} (Architect--Critic--Arbiter).}

We formalize a \textsc{PageSplit} operator by three roles, instantiate as follows. The \textbf{Architect} proposes a local candidate set $\mathcal{E}_e$ invoked through a rule-based trigger with an LLM serving as a local oracle, under either of two conditions: (i) $\mathrm{length}(e) > l_{\max}$, or (ii) a single LLM call adjudicates that $e$ admits separable entity subtrees.

The \textbf{Critic} assigns each $e \in \mathcal{E}_e$ an estimated cost change 
\begin{equation}
\widetilde{\Delta\mathcal{C}}(e; W) \;=\; \alpha\, \Delta|V| \;+\; \beta\, \Delta(\mathrm{depth}\cdot\rho) \;-\; \gamma\, \Delta\widetilde{\mathcal{Q}},
\label{eq:critic-delta}
\end{equation}
where $\widetilde{\mathcal{Q}}$ approximates $\mathcal{Q}(S; W)$ from the per-page \texttt{access\_count} and \texttt{confidence} statistics already co-located with each record (\S\ref{subsec:value-schema}).

The \textbf{Arbiter} selects a commit set $C_t \subseteq \mathcal{E}_e$
\begin{equation}
C_t \;=\; \{\,e \in \mathcal{E}_e : \widetilde{\Delta\mathcal{C}}(e; W) < 0 \,\wedge\, \mathrm{Safety}(e)\,\}, \quad |C_t| \le K,
\label{eq:arbiter-commit}
\end{equation}
where $\mathrm{Safety}(e)$ requires every entity reachable in $S_t$ to remain reachable in $S_{t+1}$, and $K$ caps the per-pass commit count.

As in \textsc{DimensionMerge}, splitting reduces per-node fan-out and lowers $\beta \sum_v \mathrm{depth}(v)\rho(v)$ for the access mass previously concentrated on $v$, at the cost of a unit increase in $|V|$ charged directly by $\alpha\,\Delta|V|$.

\paragraph*{Theorem 1 (Monotone improvement).}
Let $C$ be the schema cost~\eqref{eq:schema-cost}, bounded below. Call an operator \emph{admissible}
if it passes the acceptance test $\Delta C(e;W)\le 0$ of~\eqref{eq:arbiter-commit},
and assume the quality term $Q$ is \emph{separable across node-disjoint
supports}, so that disjoint operators contribute additively to $\Delta C$.
If each pass commits a node-disjoint set $C_t$ of admissible operators,
then $C$ is non-increasing along the greedy trajectory
$S_0\!\to\!S_1\!\to\!\cdots$ and converges.

\textit{Proof.}
The three terms of $C$ are sums over nodes, and by separability the
quality term decomposes over disjoint supports; hence for a node-disjoint
commit $\Delta C(C_t)=\sum_{e\in C_t}\Delta C(e)$, each summand
$\le 0$ by admissibility. Thus $\{C(S_t)\}$ is non-increasing and bounded
below, hence converges. $\square$

\paragraph*{Content-Level Self-Correction (Error Book).}
Complementing the schema-level operators above, our offline pipeline runs a content-level self-correction loop, the Error Book, carried over from our prior work~\cite{llmwiki}. While \textsc{DimensionMerge} and \textsc{PageSplit} operate on the structural shape of the namespace, the Error Book operates on individual record contents (e.g., dangling wikilinks, malformed source citations, unsupported facts, cross-page contradictions): detected error patterns are accumulated as constraint rules that are injected into subsequent ingestion prompts, and a two-layer repair---deterministic code-level fixes plus a periodic LLM-based fix---reduces both new and pre-existing errors. The Error Book is persisted alongside the wiki and reused across both full and incremental ingestion runs, so constraints accumulated in earlier runs continue to take effect in later ones. A contribution of this paper is to re-ground the Error Book on the storage layer, so that its constraint state and repairs share the same path-keyed records and per-author construction pipeline as the schema operators above.

\subsection{Integration with the Offline Pipeline}
\label{subsec:schema-pipeline-integration}

The cold-start procedure, the two schema-evolution operators, and the Error Book all execute as \emph{offline background jobs} that share a single construction-and-evolution pipeline with the storage layer. The pipeline runs at the following cadences: cold-start is one-shot; \textsc{DimensionMerge} and \textsc{PageSplit} are triggered every $N$ ingested articles ($N=30$ in our deployment) to absorb access-distribution drift and corpus growth; and the Error Book runs after every ingestion batch (deterministic code-level fixes) plus a periodic LLM-level fix loop, with state persisted across full and incremental runs.
\section{WikiKV Storage Model}
\label{sec:storage-model}

\begin{figure}[htbp]
    \centering
    \includegraphics[width=\columnwidth]{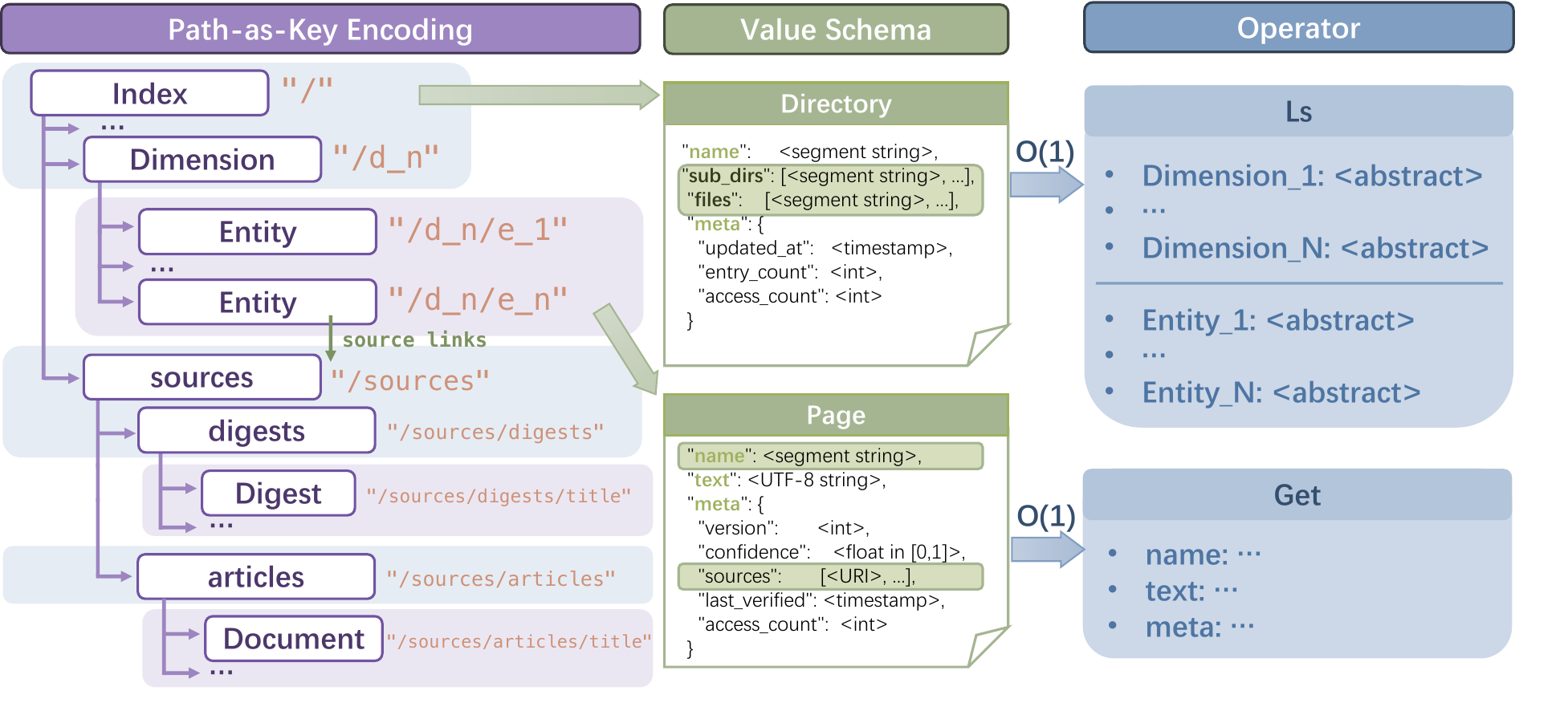}
    \caption{Path-as-Key Encoding and Value Schema in \textsc{WikiKV} (Paths are shown in logical form; the physical KV key stored in the underlying engine is the hash digest $H(\pi(v))$).}
    \label{fig:schema}
\end{figure}

We present \textsc{WikiKV}, a path-indexed storage model that materializes the wiki schema directly into the key namespace of an underlying KV store, as illustrated in Figure~\ref{fig:schema}. 

\subsection{Path-as-Key Encoding}
\label{subsec:path-as-key}

\textsc{WikiKV} encodes the wiki schema by using each node's path $\pi(v)$ \emph{verbatim} as its logical key. The five levels of the hierarchy thus collapse into a single, self-describing namespace, with the binding shown in Table~\ref{tab:path-key-binding}. The root index is bound to the literal key \texttt{"/"}. A dimension index is bound to the single-segment key \texttt{"/$d$"}. Entity is both contained by dimension directory nodes and, as a page file, contains links to original digests and documents.

Digests and documents are deliberately \emph{not} nested under each entity. Since one source article often supports several entities across dimensions, replicating its digest and full text under every referencing entity would duplicate most of the stored bytes. We therefore hoist all sources into a single shared subtree (\texttt{"/sources/digests/$\cdot$"}, \texttt{"/sources/articles/$\cdot$"}) materialized once regardless of how many entities reference them; each entity page \emph{links} to the relevant source paths rather than embedding their content, so a source shared by $k$ entities is materialized once rather than $k$ times. Consequently, corpus growth accrues to the persistent tier alone; the in-memory cache footprint is fixed by capacity caps (§ caching) and independent of corpus size.

The path $\pi(v)$ is the logical address used throughout the paper, but the \emph{physical} KV key is its hash digest $\mathrm{key}(v)=\mathrm{H}(\pi(v))$; child paths advertised in directory records (\S\ref{subsec:value-schema}) are hashed on demand. Hashing yields a fixed-width, separator- and charset-agnostic key, sidestepping the encoding pitfalls of raw path strings such as non-ASCII (e.g.\ Chinese) segments that vary in byte length and collation across backends.

\begin{table}[htbp]
\caption{Path-as-Key Binding (The physical KV key stored in the underlying engine is the hash digest $\mathrm{H}(\pi(v))$).}
\label{tab:path-key-binding}
\centering
\resizebox{\linewidth}{!}{%
\begin{tabular}{l|l|l}
\toprule
\textbf{Node} &\textbf{Logical path $\pi(v)$} & \textbf{Bound to} \\
\midrule
\textbf{Index} &\texttt{"/"}                              & root index  \\

\textbf{Dimension} &\texttt{"/$d$"}                           & dimension index $d \in V_D$ \\

\textbf{Entity} &\texttt{"/$d$/$e$"}                       & entity page $e \in V_E$ \\
\textbf{Digest} &\texttt{"/sources/digests/$title$"}                   & article digest $di \in V_{DI}$ \\
\textbf{Document} &\texttt{"/sources/articles/$title$"}                   & article document $do \in V_{DO}$ \\

\bottomrule
\end{tabular}
}
\end{table}

To make path equality unambiguous, we normalize paths: no trailing slash, case-sensitive segment matching, no reserved separator inside a segment, and depth bounded by the schema constant $D$. Normalization runs before hashing, so a path serves simultaneously as a tree address and, via $\mathrm{H}(\pi(v))$, as its storage key, with no separate translation table.

\subsection{Value Schema}
\label{subsec:value-schema}

Internal nodes (Index, Dimension) are stored as \emph{directory} records, and leaves (Entity, Document and Digest) are stored as \emph{file} records.

\paragraph*{Directory records.}
A directory record has three parts. It identifies the node via \texttt{type}$=$\texttt{"dir"} and a \texttt{name} holding its segment relative to the parent; it enumerates children with two parallel arrays, \texttt{sub\_dirs} (child directories) and \texttt{files} (child leaves); and it carries a \texttt{meta} block of statistics: \texttt{updated\_at}, \texttt{entry\_count}, and an \texttt{access\_count}.

\paragraph*{File records.}
A file record follows an analogous layout: \texttt{type}$=$\texttt{"file"} with a \texttt{name}, a single UTF-8 \texttt{text} payload, and a \texttt{meta} block recording a monotonically increasing \texttt{version}, a \texttt{confidence} score in $[0,1]$, a \texttt{sources} list of URIs, a \texttt{last\_verified} timestamp, and an \texttt{access\_count}.

\paragraph*{Design rationale.}
Two points deserve emphasis. First, since every directory record names its reachable children explicitly, $\textsc{Ls}(\pi)$ is served by a single point lookup at $\pi$ with no prefix scan or auxiliary index, i.e.\ $\textsc{Ls}(\pi)\equiv\textsc{Get}(\pi)$ in $O(1)$ time in the common case. Second, the \texttt{meta} counters (\texttt{access\_count}, \texttt{confidence}, \texttt{last\_verified}, \texttt{version}) are unused by the storage operators here but feed the schema-evolution operators of Section~\ref{sec:schema-evolution}. 

\subsection{Consistency Protocol}
\label{subsec:consistency-protocol}

\begin{figure}[htbp]
    \centering
    \includegraphics[width=\columnwidth]{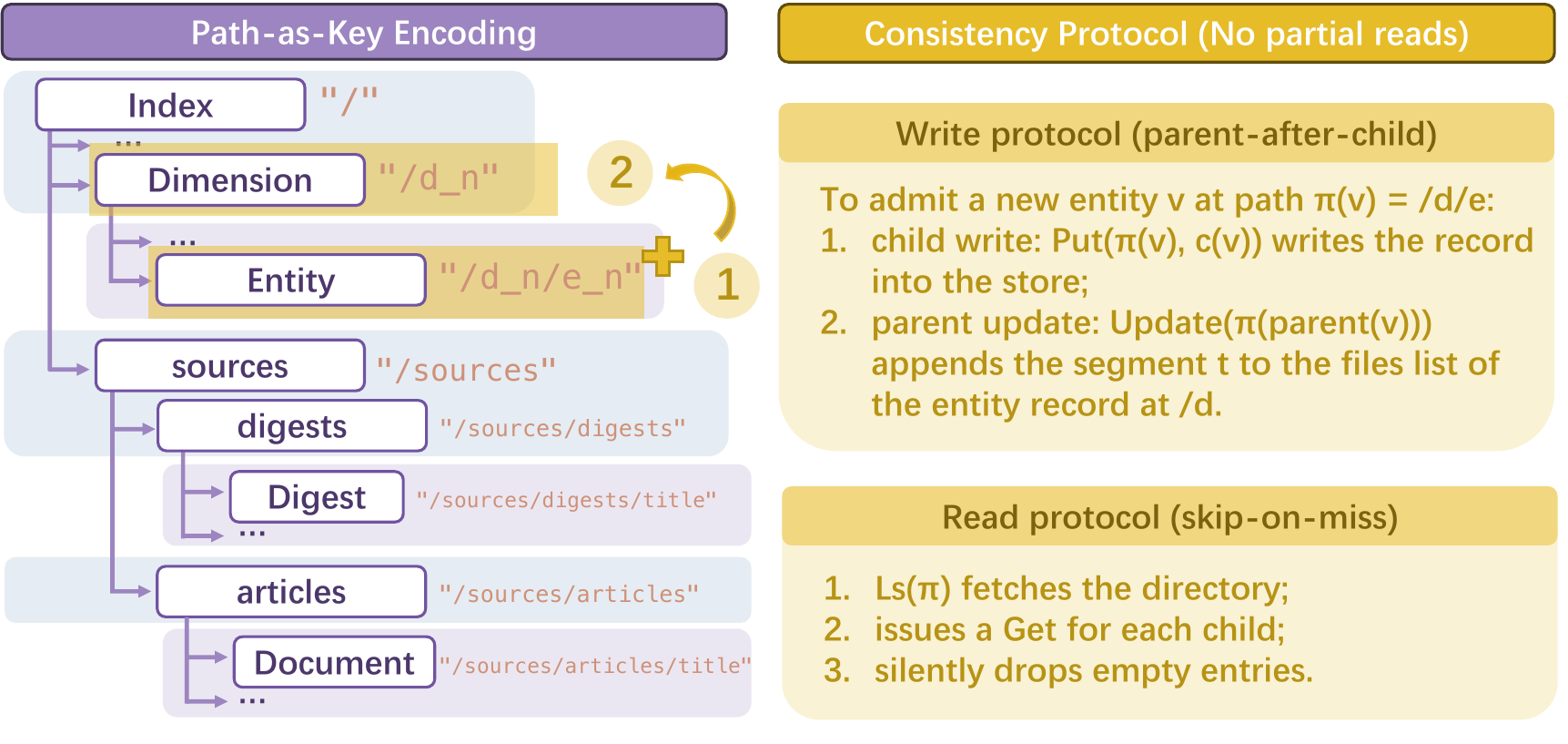}
    \caption{Consistency Protocol(No partial reads): Write protocol (parent-after-child) and Read protocol (skip-on-miss).}
    \label{fig:protocol}
\end{figure}

The storage layer must reconcile two requirements: newly admitted pages must be reachable through their parent directory (R1), yet readers running concurrently with offline writes must never observe a partial-write state (R2). As shown in Figure~\ref{fig:protocol}, \textsc{WikiKV} discharges both with a parent-after-child write protocol paired with a permissive read, taking no explicit locks on the read path.

\paragraph*{Write protocol (parent-after-child).}
To admit a new entity $v$ at path $\pi(v) = /d/e$, the offline pipeline executes the following two operations \emph{in order}: (1) \textbf{Child write.} $\textsc{Put}(\pi(v), c(v))$ writes the entity record into the store. (2) \textbf{Parent update.} $\textsc{Update}(\pi(\mathrm{parent}(v)))$ appends the segment $e$ to the \texttt{files} list of the dimension record at $/d$.

If step~2 fails, the orphan record at $\pi(v)$ remains in the store but is not yet linked from any directory listing. Page-level updates that rewrite an existing entity in place follow the same pattern, except that step~2 typically only refreshes \texttt{meta} statistics on the parent and is therefore a no-op.

\paragraph*{Read protocol (skip-on-miss).}
A reader executing $\textsc{Ls}(\pi)$ first fetches the directory record at $\pi$ and then issues a $\textsc{Get}$ for each child path it advertises. If a child $\textsc{Get}$ returns $\bot$, the reader silently drops that entry from the result. Combined with the write order above, this discipline rules out partial-write observations, as the next theorem makes precise.

\paragraph*{Theorem 2 (No partial reads).}
\label{thm:no-partial-read}
Under the parent-after-child write and skip-on-miss read protocols, and
assuming monotonic cross-key visibility (a reader observing the parent
update also observes the prior child write), a \textsc{Get}/\textsc{Ls} on
$\pi$ never returns a child that the record at $\pi$ advertises but whose
own record is missing.

\textit{Proof.} Let a reader observe a directory record $R$ at $\pi$ and $\pi(v)$ be a new
entity. (a) If $R$ omits $\pi(v)$, step~2 has not committed, so $v$ is at
most an unadvertised orphan and is not fetched. (b) If $R$ lists it, step~2
committed, so by the write order and cross-key visibility step~1 is durable
and visible here, and $\textsc{Get}(\pi(v))$ is complete. The only orphan
reachable via a stale cache listing returns $\bot$ and is dropped by
skip-on-miss before reaching the result set. $\square$

This guarantee is stated for the deployment we evaluate---a single offline writer per subtree with a read-only online tier---and concerns advertised-but-missing children rather than full snapshot isolation; concurrency across multiple offline writers and strong isolation under cache races are delegated to the underlying \textsc{TableKV} layer.

\paragraph*{Optimistic concurrency control.}
Page-level updates rewrite existing records, so we attach a monotonically increasing \texttt{version} field to every record and use it as a compare-and-swap token. A writer that finds its expected version stale aborts and retries with the latest value. Because the workload is read-only on the online tier and the offline pipeline is the sole writer, write--write conflicts are rare in practice and a small bounded number of retries suffices; we therefore do not require a stronger pessimistic locking layer.

\paragraph*{Multi-process parallel construction.}
The offline pipeline scales out by \emph{partitioning the write workload along author boundaries}: each author's corpus compiles into its own subtree, and distinct authors share no path, so their write sets are disjoint by construction. Construction is thus \emph{per-author-parallel, intra-author-serial}---a pool of workers each owns one author at a time and applies its ingestion and evolution steps in sequence (preserving the parent-after-child order and version-CAS within the subtree), while different workers proceed in parallel with no cross-author coordination. This introduces no write--write conflicts beyond the intra-subtree ones already handled by OCC, and Theorem~2 continues to hold per subtree. With each ingestion batch completing in a few seconds, a modest worker pool sustains the throughput needed to build and maintain wikis for millions of Official Account authors.

\ifcsname c@algorithm\endcsname\else\newcounter{algorithm}\fi

\section{Budgeted Path Navigation Query}
\label{sec:navigation-query}

Building on the key--value primitives of the storage layer, this section introduces the budgeted navigation query operator $\textsc{Nav}(q, B)$, which compiles a multi-step descent into $O(1)$ search-accelerated, LLM-assisted hops and pairs it with a path-keyed three-tier cache to sustain stable tail latency across knowledge-base scales.

\subsection{Query Semantics}
\label{subsec:query-semantics}

The navigation query operator $\textsc{Nav}(q, B)$ is the only operator in \S\ref{subsec:query-model} not mapped to a single storage primitive: given a natural-language query $q$ and wall-clock budget $B$, it returns an \emph{ordered} sequence $R = \langle r_1, \ldots, r_m \rangle$ with $m \le \lceil B/b \rceil$, where $b$ is the dominant single-step latency (an LLM-assisted descent in the worst case, a $\textsc{Get}$ in the best case). The operator is thus \emph{anytime}: increasing $B$ may yield a longer, finer sequence without invalidating the prefix returned under a smaller budget.

\begin{figure}[htbp]
    \centering
    \includegraphics[width=\columnwidth]{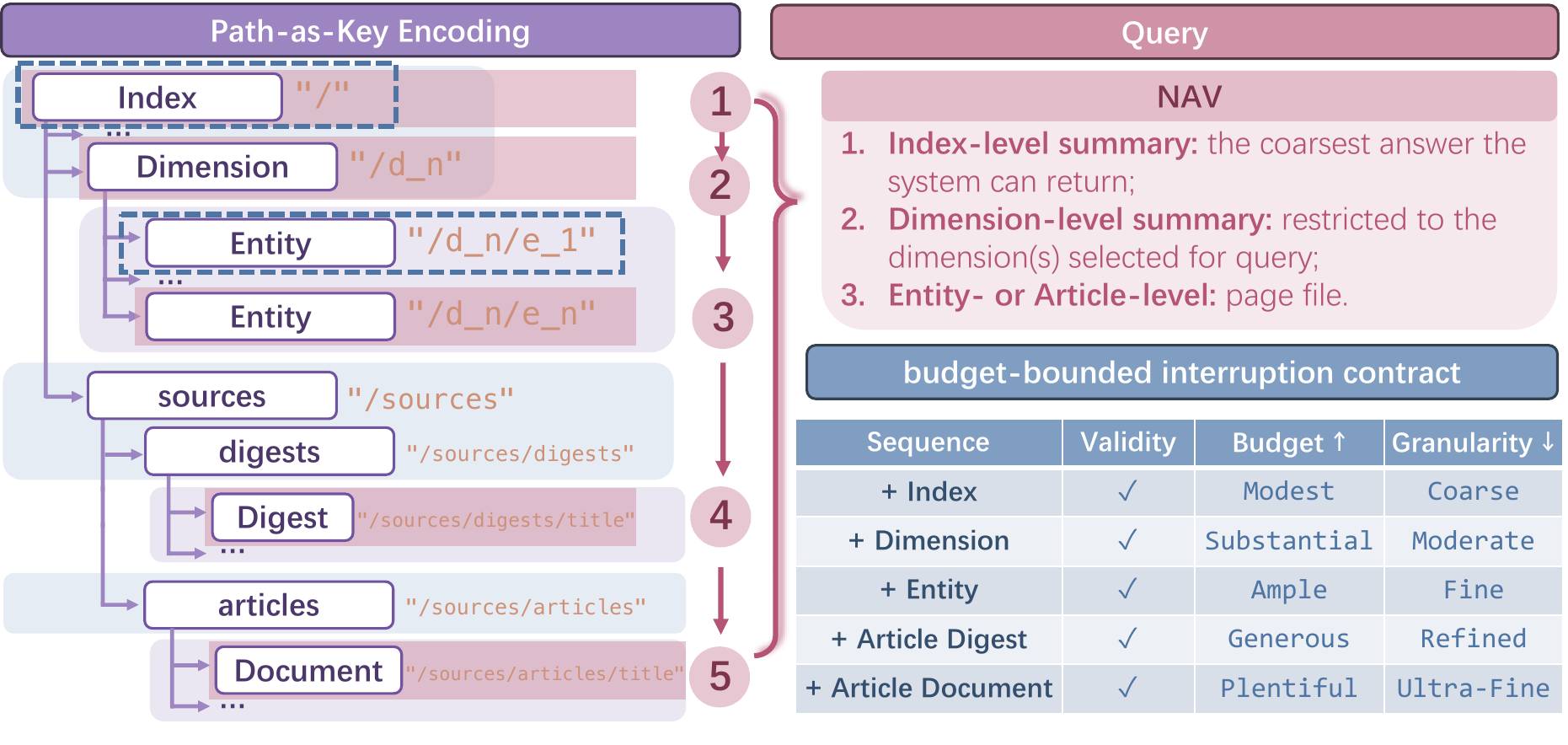}
    \caption{Flowchart of the navigation query operator with budget-bounded interruption contract.}
    \label{fig:navi}
\end{figure}
As illustrated in Figure \ref{fig:navi}, $\textsc{Nav}$ differs from a generic anytime traversal in the \emph{progressive} contract of Property~1 (\S\ref{subsec:query-model}), specialized here to the wiki schema: records are emitted in order of monotonically increasing granularity, aligned with the hierarchy levels:
\begin{itemize}
  \item $r_1$ is an \emph{Index-level summary}, the coarsest answer the system can return; e.g., ``the wiki contains $N$ dimensions: \textit{Personal Relationships}, \textit{Writing Style}, \ldots''.
  \item $r_2$ is a \emph{Dimension-level summary} restricted to the dimension(s) selected for $q$; e.g., ``\textit{Personal Relationships} contains three entities: \textit{Family}, \textit{Mentors and Friends}, \textit{Polemic Opponents}''.
  \item $r_3$ and onward are \emph{Entity- or Article-level} pages; e.g., ``The estrangement between Zhou Zuoren and Lu Xun occurred in 1923 \ldots''.
\end{itemize}
For any prefix length $i \le m$, $\langle r_1, \ldots, r_i \rangle$ is a valid (if coarser) answer to $q$. The operator thus admits a \emph{budget-bounded interruption} contract: when the budget runs out, the accumulated partial sequence is returned as-is. This motivates the emission order of the plan and the budget-checking guards in the algorithm.

\subsection{Search-Accelerated Routing}
\label{subsec:search-routing}

A literal layer-by-layer execution of $\textsc{Nav}(q, B)$ would invoke an LLM at each level to pick the child to descend, incurring $D$ serial LLM calls before the first entity-level record. The path-as-key namespace short-circuits this descent: a lexical keyword/prefix search $\textsc{Search}(p)$ over the path namespace (\S\ref{subsec:query-model}) returns candidate target paths that already approximate the right region of the tree, with no per-level LLM call. The router operates on \emph{textual path keys}, not a dense vector index; vector retrieval, where used, is confined to the leaf-content operator Q4 and is orthogonal to path routing. The plan thus has two phases: Phase~1 selects $k$ candidate paths via a single search call (a constant number of KV round trips, independent of $D$); Phase~2 performs targeted point lookups and optional single-level expansions. Both phases are gated by budget checks, so the operator can return at any point with a coherent prefix.

\paragraph*{Engineering components.}
The algorithm comprises five lightweight components, each an explicit pseudocode step rather than an opaque sub-procedure.
\begin{itemize}
  \item \textsc{Classify}$(q)$ is a hybrid router combining a regular-expression layer for enumeration triggers (``which \ldots'', ``list \ldots'') with a small distilled classifier for ambiguous queries, adding at most $5$ms. Its route class $\textit{cls}$ drives Phase~1: enumeration queries are answered by a single directory listing, the rest forwarded to the search router.
  \item \textsc{Search}$(\textsc{Extract}(q))$ extracts candidate page-name keywords from $q$ and runs a prefix scan over the path namespace to return candidate file paths.
  \item \textsc{NeedsDeeper}$(q, v)$ compares $q$ against a candidate's content $c(v)$ and returns \texttt{true} only if its semantic coverage of $q$ falls below a threshold $\theta$; it is a lightweight classifier or a single LLM call.
  \item \textsc{BudgetExhausted}$(t_0, B)$ is the budget enforcement guard, evaluated before every potentially expensive step.
  \item Multi-path exploration over the candidate set is performed serially by default to keep the budget guard authoritative; when the residual budget is large enough to amortize the call overhead, the per-candidate $\textsc{Get}$ calls of Phase~2 may be issued in parallel.
\end{itemize}

\paragraph*{Algorithm.}
Algorithm~\ref{alg:nav} states the plan, with $t_0$ the start time, $\textit{cls}$ the route class from \textsc{Classify}, $\Pi$ the Phase~1 candidate set, and $R$ the accumulating result. Enumeration-style queries short-circuit to a single root listing, bypassing keyword search; all other classes proceed to search-accelerated routing. Lines~\ref{alg:nav:guard1} and~\ref{alg:nav:guard2} are the budget guards, and line~\ref{alg:nav:expand} performs the optional single-level expansion when a candidate page does not by itself cover $q$.

\begin{figure}[htbp]
\hrule\vspace{4pt}
\textbf{Algorithm~1: \textsc{Nav}$(q, B)$: Search-Accelerated Navigation}
\refstepcounter{algorithm}\label{alg:nav}
\vspace{2pt}\hrule\vspace{4pt}
\begin{algorithmic}[1]
\REQUIRE Natural-language query $q$, time budget $B$ (ms)
\ENSURE Progressive result sequence $R$
\STATE $t_0 \gets \textsc{Now}()$;\quad $R \gets \langle\,\rangle$
\STATE $\textit{cls} \gets \textsc{Classify}(q)$ \COMMENT{$\le 5$ms hybrid router}
\STATE \emph{// Phase 1: search-accelerated routing}
\IF{$\textit{cls} = \textsc{Enumerate}$}
  \STATE \textbf{return} $\langle \textsc{Ls}(\texttt{"/"})\rangle$ \COMMENT{enumeration query answered by a single directory listing}
\ENDIF
\STATE $\Pi \gets \textsc{TopK}(\textsc{Search}(\textsc{Extract}(q)), k)$ \COMMENT{$k\!=\!3$ candidate paths}
\IF{\textsc{BudgetExhausted}$(t_0, B)$} \label{alg:nav:guard1}
  \STATE \textbf{return} $\langle \textsc{Ls}(\texttt{"/"})\rangle$ \COMMENT{coarsest fallback}
\ENDIF
\STATE \emph{// Phase 2: targeted navigation}
\FORALL{$\pi \in \Pi$}
  \STATE $v \gets \textsc{Get}(\pi)$;\quad $R.\textsc{Append}(v)$
  \IF{$\textsc{NeedsDeeper}(q, v)$}
    \STATE $R.\textsc{Append}(\textsc{Ls}(\pi))$ \label{alg:nav:expand}
  \ENDIF
  \IF{\textsc{BudgetExhausted}$(t_0, B)$} \label{alg:nav:guard2}
    \STATE \textbf{break}
  \ENDIF
\ENDFOR
\STATE \textbf{return} $R$
\end{algorithmic}
\vspace{2pt}\hrule
\end{figure}

\paragraph*{Theorem 3 (Step compression).}
Let $D$ be the wiki depth and let $h$ denote the number of post-routing hops needed to reach a target file from a candidate path returned by Phase~1. Under Algorithm~1, the number of LLM-assisted descent steps along a single target path drops from $D$ in pure layer-by-layer navigation to $h$, where $h \in \{0,1\}$ for single-target queries (the candidate already covers $q$, or one expansion via \textsc{NeedsDeeper} suffices) and $h \le k$ when $q$ requires aggregating evidence across $k$ dimensions.

\textit{Proof.} Layer-by-layer navigation needs one LLM call per level, i.e.\ $D$ to reach
depth $D$. Under Algorithm~1, Phase~1 issues one routing call, then a single
\textsc{Search} replaces the first $D-h$ levels in constant KV round-trips
independent of $D$. Among the rest only \textsc{NeedsDeeper} is on the
critical path, giving the $h$ post-routing descent calls: zero if the
candidate covers $q$, once if one \textsc{Ls} expansion suffices, at most
$k$ when $q$ spans $k$ dimensions. Thus the descent steps equal $h$ (the
quantity in the statement) and the end-to-end count is the routing call plus
these, $1+h$; both are independent of $D$. $\square$

\subsection{Caching Strategy}
\label{subsec:caching-strategy}

The query model of \S\ref{subsec:query-model} treats every $\textsc{Get}$ as a fixed KV round trip, but in production we observe a strongly skewed access distribution: the root index and a small handful of dimension pages are read on essentially every query, while deep entity pages are read rarely but must remain available for the budgeted descent. This skew motivates a three-tier cache hierarchy keyed on the same path namespace as the underlying store, so that the cache lookup and the storage lookup share a single key derivation. The three tiers are arranged in increasing order of capacity and decreasing order of expected hit rate.

\paragraph*{L1 --- in-process cache (capacity: tens of pages).}
\begin{itemize}
  \item \emph{Contents.} The root index node \texttt{"/"} and every dimension node \texttt{"/$d$"}.
  \item \emph{Policy.} Pre-warmed at process start, never expired during the lifetime of the process; refreshed on the cache-invalidation event described below.
  \item \emph{Rationale.} Bounds the worst-case latency of Phase~1 by removing the single most frequently accessed prefix of the namespace from the network path.
\end{itemize}

\paragraph*{L2 --- shared Redis tier (capacity: thousands of pages).}
\begin{itemize}
  \item \emph{Contents.} Full directory node plus hot subset of Entity nodes, identified by the \texttt{access\_count} statistic kept in each file's \texttt{meta} record (\S\ref{subsec:value-schema}).
  \item \emph{Policy.} LRU eviction with a one-hour TTL, so that pages displaced from the working set are eventually reclaimed even without an explicit invalidation.
  \item \emph{Rationale.} Absorbs the long tail of dimension- and entity-level traffic that L1 cannot fit, and serves as the cross-process cache shared by all online query workers.
\end{itemize}

\paragraph*{L3 --- KV store (capacity: full wiki).}
\begin{itemize}
  \item \emph{Contents.} The complete \textsc{WikiKV} keyspace, including full directory and file nodes.
  \item \emph{Policy.} Persistent backing tier; serves any request that misses L1 and L2.
  \item \emph{Rationale.} Acts as the authoritative source of truth and the consistency anchor for the cache-invalidation protocol.
  
\end{itemize}

\paragraph*{Bounded in-memory footprint.} Only L1 and L2 reside in memory, and both are capped by a fixed page budget (tens and thousands of pages, respectively) with LRU+TTL eviction; the resident working set is therefore bounded by the access skew rather than by corpus size. Growth in the article count enlarges only the persistent L3 tier (disk/SSD), so memory usage stays flat as the wiki scales. We deliberately attach no expiration to L3: a curated knowledge base is authoritative and durable, and content staleness is handled actively by the cache-invalidation stream and the Error Book / schema-evolution operators, not by passively expiring stored knowledge.
\paragraph*{Cache invalidation.}
The offline construction-and-evolution pipeline publishes a path-keyed invalidation event on every successful write that completes the parent-after-child protocol of \S\ref{subsec:consistency-protocol}. The online tier subscribes to this event stream and asynchronously refreshes any L1 or L2 entry whose key is a prefix of, or equal to, the affected path. Because the parent-after-child protocol guarantees that no advertised-but-missing child is ever observable in the underlying store (Theorem~2), an invalidation that races with an in-flight read can at worst force an extra round trip to L3, but it can never expose a partial-write state to the application. The bounded-staleness requirement R3 is therefore satisfied with $\Delta$ equal to the maximum delay between a successful offline commit and the corresponding cache-refresh callback, which is independent of the wiki size.

\section{Experimental Evaluation}
\label{sec:experiments}

This section evaluates \textsc{WikiKV} on a production hierarchical knowledge-base platform, against both representative storage backends and representative retrieval-augmented baselines. 

\subsection{Setup}
\label{subsec:setup}

\paragraph*{Dataset and workload.}
We use \textsc{AuthTrace}~\cite{authtrace2026}, a recent diagnostic benchmark for evidence construction over thematically dense single-author corpora. It provides quoted evidence, exact fan-in annotations per question, and a unified pack-level protocol that measures evidence recall, evidence precision, and answer correctness. Its \emph{fan-in gradient}, i.e., the number of source documents required to support an answer, defines a primary diagnostic axis along which retrieval, memory, graph, and structured-evidence paradigms can be compared on a level playing field. We adopt its three buckets: \emph{single-doc} (evidence in one document), \emph{low multi-doc} (two documents), and \emph{high multi-doc} (three or more documents).

\paragraph*{Platform.}
The online tier serves user-facing navigation queries, while an offline construction-and-evolution pipeline writes pages to the storage layer. The read and write paths share no synchronous coordination beyond the parent-after-child write protocol of \S\ref{subsec:consistency-protocol}.
The online tier runs on a Redis Cluster and the persistent tier on \textsc{TableKV}, a distributed KV system built by WeChat on an LSM-tree engine with PaxosStore consensus, whose simple KV abstraction scales horizontally. All LLM inference uses \textsc{DeepSeek-V4-Flash}\cite{deepseekai2026deepseekv4}.\footnote{\url{https://huggingface.co/deepseek-ai/DeepSeek-V4-Flash}}

\paragraph*{Baselines.}
The storage latency study (\S\ref{subsec:microbenchmarks}) compares \textsc{WikiKV}'s path-as-key layout on its LevelDB engine against 
three alternative backends: a hierarchical file system (FS) using 
directory and file primitives; PostgreSQL with the ltree extension 
and a normalized parent-child schema; and Neo4j, representing native 
property-graph storage. The end-to-end study (\S\ref{subsec:end-to-end}) compares \textsc{WikiKV} against retrieval-augmented baselines: \emph{No-RAG}, which queries the LLM with no retrieved evidence; \emph{Dense-RAG}, embedding-based retrieval over a flat vector index; \emph{GraphRAG}~\cite{edge2024graphrag}, community-summary retrieval over a constructed knowledge graph; and \emph{RAPTOR}~\cite{sarthi2024raptor}, recursive abstractive summarization producing a hierarchical retrieval tree.
\paragraph*{Measurement protocol.}
All latency numbers are medians over 10 independent runs after a
200-query warmup; inter-run variation stays below 1\% throughout.
To control for LLM stochasticity, every answer-correctness result
uses greedy decoding (temperature $0$) with a fixed seed, so the
reported scores are deterministic given the retrieved evidence.
Unless noted otherwise, accuracy gaps reported below are
statistically significant under a paired bootstrap test over
per-question scores ($p < 0.01$).

\subsection{Storage Backend Latency}
\label{subsec:microbenchmarks}

We first compare per-operator latency corresponding to Q1--Q4 in (\S\ref{subsec:query-model}) across the four storage configurations using a medium-sized wiki ($\sim$2{,}000 KV pairs), as shown in Table~\ref{tab:microbench-p50}.
For each operator, we randomly sample 100 target paths or prefixes and issue 1{,}000 queries per backend after a 200-query warmup.
The path-as-key engine is our method: to isolate engine cost from service-layer effects, we realize it on a local LevelDB exposing the same Put/Get interface as \textsc{TableKV} but bypassing the service stack, reflecting only the local engine's read/write cost. The remaining three backends—FS, PostgreSQL, and Neo4j—are the comparison points. All backends use a controlled, in-process, memory-resident setup; absolute latencies are therefore lower than in production, but since every backend is measured under the same configuration the relative comparison remains fair.

\begin{table}[htbp]
\caption{Median (P50) Per-Operator Latency by Backend, in Milliseconds, on the \textsc{Medium} Wiki.}
\label{tab:microbench-p50}
\centering
\renewcommand{\arraystretch}{1.15}
\begin{tabular}{l|cccc}
\toprule
\textbf{Backend} & \textbf{Q1} & \textbf{Q2} & \textbf{Q3} & \textbf{Q4} \\
\midrule
FS                & 0.021 & 0.436 & 6.480 & 0.091 \\
\cellcolor{gray!12}\textsc{WikiKV}           & \cellcolor{gray!12}0.017 & \cellcolor{gray!12}0.088 & \cellcolor{gray!12}5.411 & \cellcolor{gray!12}0.085 \\
PostgreSQL        & 0.075 & 0.117 & 1.671 & 0.432 \\
Neo4j             & 2.494 & 1.230 & 6.397 & 1.686 \\
\bottomrule
\end{tabular}

\end{table}

Our claim is therefore not that a path-as-key engine is the single fastest backend on every operator---none dominates on all four---but that it attains \emph{balanced} low latency \emph{across} the full Q1--Q4 mix, whereas each alternative is fast on some operators and pays disproportionately on others.
\paragraph*{Q1 (path lookup).}
\textsc{WikiKV} and FS deliver sub-millisecond P50 because Q1 maps directly to a single-key fetch in either backend. PostgreSQL pays a small but consistent overhead from the SQL parsing path even though the \texttt{ltree} key is indexed, while Neo4j is two orders of magnitude slower than \textsc{WikiKV} because every Q1 incurs Bolt-driver round-tripping and a Cypher plan compilation. Backends whose data path is closest to the path-as-key contract pay the lowest constants.

\paragraph*{Q2 (directory listing).}
\textsc{WikiKV} is the fastest at P50 because the value record co-locates child segments (\S\ref{subsec:value-schema}) so that Q2 reduces to a single point lookup rather than a prefix scan. FS is slowed by per-entry metadata syscalls, and PostgreSQL is competitive only because the \textsc{Medium} working set fits in memory and the \texttt{ltree} index avoids a recursive CTE. Neo4j again pays a multi-millisecond constant because the operation must traverse outgoing edges and rebuild Cypher result rows.

\paragraph*{Q3 (navigation along a known path).}
Q3 is where the four backends diverge most clearly. PostgreSQL is unexpectedly the fastest at P50 (1.671\,ms) because the simulated navigation decomposes into a small number of indexed path equality lookups, each of which reuses a single client connection; with only $\sim$2{,}000 rows the index and base table are essentially in the page cache, and the per-step SQL constant is small. \textsc{WikiKV} and FS both incur multiple round trips through their respective directory layers (a prefix scan and JSON parse for \textsc{WikiKV}, repeated metadata calls for FS), and Neo4j additionally pays the Bolt and Cypher constants once per descent step. 

\paragraph*{Q4 (prefix search).}
\textsc{WikiKV} and FS are the strongest at P50 because their lexicographic key layouts permit a native prefix range scan. PostgreSQL incurs additional overhead through the \texttt{ltree} match operator, and Neo4j has no native prefix-search primitive---it emulates Q4 with a pattern-match query that requires plan compilation per call.

\subsection{Schema Design and Evolution Effectiveness}
\label{subsec:schema-evolution-eval}

We next isolate the contribution of the cold-start procedure (\S\ref{subsec:schema-cold-start}) and the two schema-evolution operators (\S\ref{subsec:evolution-operators}). The full \textsc{WikiKV} system is compared against two internal variants on \textsc{AuthTrace}: \textsc{WikiKV-FixedSchema} replaces cold-start with fixed dimensions, and \textsc{WikiKV-Static} keeps the cold-started schema but disables both evolution operators. All three configurations share the same storage layer (\S\ref{sec:storage-model}) and query layer (\S\ref{sec:navigation-query}), so any difference in answer correctness or latency is attributable to schema design and evolution alone. Throughout this section, \emph{answer correctness} (AC) denotes the end-to-end correctness of the generated answer under the \textsc{AuthTrace} pack-level protocol; we report AC together with online \emph{first-token latency}, and reserve \emph{human rating} for the production study of \S\ref{subsec:production}.

\begin{table}[htbp]
\caption{Effect of Cold-Start and Evolution on Answer Correctness and Online Latency}
\label{tab:schema-evolution}
\centering
\renewcommand{\arraystretch}{1.15}
\begin{tabular}{l|c|c|c}
\toprule
\textbf{Metric} & \textbf{\textsc{WikiKV}} & \textbf{\textsc{Fixed}} & \textbf{\textsc{Static}} \\
\midrule
Page count       & 2385           & 2672  & 2290  \\
Avg.\ tool calls / query     & 3.18 & 3.36 & 3.12 \\
Avg.\ pages read / query     & 1.48 & 1.56 & 1.44 \\
\midrule
Avg.\ first-token time (s)($\downarrow$)    & \textbf{11.7} & 12.7 & \textbf{11.7} \\

AC ($\uparrow$)              & \textbf{63.2} & 53.5 & 52.5 \\
\bottomrule
\end{tabular}

\end{table}

\paragraph*{Cold-start vs.\ fixed dimensions.}
Replacing the cold-start procedure with a manually fixed schema (\textsc{Fixed}) costs nearly ten absolute points of answer correctness (53.5 vs.\ 63.2), and simultaneously inflates the page count by 12\% (2672 vs.\ 2385) because fixed dimensions over-partition the corpus into thematically thin subtrees. The accuracy loss propagates to online latency---average first-token time rises from 11.7\,s to 12.7\,s---because the wider Index and the over-partitioned topics force the navigation operator into more \textsc{NeedsDeeper} expansions per query (3.36 vs.\ 3.18 average tool calls) and into reading more pages per descent (1.56 vs.\ 1.48). The data-driven cold-start of \S\ref{subsec:schema-cold-start} therefore dominates fixed dimensions across every metric we report.

\paragraph*{Evolution operators vs.\ frozen catalog.}
Freezing the schema after cold-start (\textsc{Static}) preserves the latency profile of full \textsc{WikiKV}---average first-token time is identical to the full system (11.7\,s)---but loses more than ten absolute points of answer correctness (52.5 vs.\ 63.2). The page count (2290 vs.\ 2385) further reveals the mechanism: without \textsc{PageSplit} and \textsc{DimensionMerge}, the schema fails to grow new topics in regions where the corpus has densified and fails to coalesce dimensions that the access distribution treats as one concept. The two evolution operators of \S\ref{subsec:evolution-operators} therefore contribute the bulk of the system's accuracy headroom while imposing essentially no latency overhead, consistent with Theorem~1's monotone-improvement guarantee.

\paragraph*{Page-count and accuracy interaction.}
Read jointly, the two ablations in Table~\ref{tab:schema-evolution} establish that schema quality, not schema size, is the binding constraint. \textsc{Fixed} is the largest schema yet the least accurate, while \textsc{Static} is the smallest yet still loses ten points to full \textsc{WikiKV}; only the cold-start--plus--evolution combination achieves both an appropriately sized namespace and the highest accuracy. This is the empirical analogue of the joint optimization in Eq.~\ref{eq:schema-cost}: minimizing $|V|$ alone (\textsc{Static}) or maximizing fan-out alone (\textsc{Fixed}) is insufficient; both terms must be co-optimized through the operators of \S\ref{subsec:evolution-operators}.

\subsection{End-to-End Retrieval: AuthTrace}
\label{subsec:end-to-end}

We compare the inherited \textsc{LLM-Wiki} retrieval pipeline of \S\ref{sec:navigation-query} running on top of \textsc{WikiKV} against four representative baselines on \textsc{AuthTrace}.
All baselines share the same generation model and prompt template; only the retrieval stage differs. The metric is answer correctness, reported per fan-in bucket and as an overall average (Table \ref{tab:end-to-end}).

\begin{table}[htbp]
\caption{End-to-End Answer Correctness on \textsc{AuthTrace} by Fan-in Bucket}
\label{tab:end-to-end}
\centering
\renewcommand{\arraystretch}{1.15}
\resizebox{\linewidth}{!}{%
\begin{tabular}{l|c|c|c|c}
\toprule
\textbf{Method} & \textbf{Single-doc} & \textbf{Low multi-doc} & \textbf{High multi-doc} & \textbf{Overall} \\
\midrule
No-RAG          & 9.9                 & 16.4                   & 16.1                    & 12.4 \\
Dense-RAG       & 59.5                & 37.1                   & 29.4                    & 49.9 \\
GraphRAG        & 52.2                & 33.8                   & 27.6                    & 44.3 \\
RAPTOR          & 60.1                & 35.6                   & 30.8                    & 50.0 \\
\cellcolor{gray!12}\textbf{LLM-Wiki (\textsc{WikiKV})} & \cellcolor{gray!12}\textbf{67.2} & \cellcolor{gray!12}\textbf{60.9} & \cellcolor{gray!12}\textbf{47.7} & \cellcolor{gray!12}\textbf{63.2} \\
\bottomrule
\end{tabular}
}

\end{table}

\paragraph*{Single-doc bucket.}
On single-doc queries, \textsc{WikiKV} leads by 7.1 points ($p<0.01$) over the strongest baseline (RAPTOR). Single-doc questions are precisely the regime in which a flat vector index is expected to perform best, because a single near-neighbor lookup can already locate the supporting passage; the gap above Dense-RAG and RAPTOR is therefore attributable to the structural advantage of navigating to a path-bounded entity page rather than to a free-floating chunk.

\paragraph*{Low-multi-doc bucket.}
The gap widens dramatically at fan-in 2: \textsc{WikiKV} delivers 60.9 versus the strongest baseline's 37.1 (Dense-RAG). Two-document evidence requires either explicit traversal between sibling entity pages or multi-level traversal. \textsc{WikiKV}'s path-indexed navigation matches this access pattern directly, while flat dense retrieval and graph-community retrieval both lack such capability.

\paragraph*{High-multi-doc bucket.}
On the highest-fan-in bucket, \textsc{WikiKV} scores 47.7 against the best baseline's 30.8, a 16.9-point gap (paired bootstrap, $p < 0.01$) that confirms the prediction of \S\ref{subsec:search-routing}: as the number of evidence documents required by a query grows, the value of structural navigation grows correspondingly, because each additional supporting document is typically reachable via hierarchical wiki nodes. The cross-bucket trend---the three retrieval baselines lose 20+ points moving from single-doc to high-multi-doc, while \textsc{WikiKV} loses fewer than 20---indicates that structural retrieval degrades more gracefully under fan-in stress than any of the flat-index or summary-tree baselines.

\subsection{Production Deployment Study}
\label{subsec:production}

We complement the above with a live deployment of \textsc{WikiKV}
as a personal knowledge-base service for the WeChat Official Account
AI Assistant. The deployment backs a production knowledge base on
the order of millions of pages and tens of millions of KV pairs
across hundreds of author accounts, sustaining query traffic several
orders of magnitude above the controlled workload of
\S\ref{subsec:end-to-end}.%
\footnote{Absolute corpus and traffic figures are withheld for
commercial confidentiality; the reported orders of magnitude suffice
to characterize the deployment scale.}
We sample 1{,}000 real user queries against author-specific knowledge
bases and measure the full online path (router $\to$ navigation $\to$
generation), reporting both system-side latency and human-evaluated
quality.

\paragraph*{Quality-grading protocol.}
The 1{,}000 queries are uniformly sampled from the live query log and de-identified before review. Each answer is rated on a 3-point scale that follows the production review rubric: \textbf{3 (exact hit)}---the answer directly answers, or fully supports answering, the user query without additional inference or supplementary information; \textbf{2 (related but indirect)}---the answer does not directly answer the query but contains content thematically related to the query that can be used after further inference; \textbf{1 (irrelevant or empty)}---the answer has no substantive connection to the query, or is empty. Each item is graded by three annotators blind to the system configuration, with substantial inter-annotator agreement (Krippendorff's $\alpha = 0.71$); the reported value is the per-item mean.

\begin{table}[htbp]
\caption{Online Latency and Human-Evaluated Quality on 1{,}000 Real Queries from the WeChat Official Account AI Assistant Deployment}
\label{tab:production}
\centering
\renewcommand{\arraystretch}{1.15}
\begin{tabular}{l|c|c|c|c}
\toprule
\textbf{Metric} & \textbf{Avg.} & \textbf{P50} & \textbf{P95} & \textbf{P99} \\
\midrule
Wiki tool calls / query       & 2.2   & 2      & 3      & 4.6     \\
Wiki tool latency (s)         & 0.432 & 0.411  & 0.554  & 0.966  \\

First-token latency (s)       & 6.856 & 6.65  & 8.34 & 12.32 \\

\midrule
Human rating (1--3)           & \multicolumn{4}{c}{2.86} \\
\bottomrule
\end{tabular}

\end{table}

\paragraph*{Online latency (Table~\ref{tab:production}).}
The deployed system delivers a mean first-token latency of 6.856\,s. The wiki-tool component itself contributes only 0.432\,s on average and remains under 1\,s even at P99, confirming that the path-as-key storage path of \S\ref{sec:storage-model} is not the bottleneck in production: end-to-end time is dominated by LLM generation rather than by retrieval. The mean tool-call count of 2.2 per query (median 2, P99 4.6) further indicates that the search-routing operator of \S\ref{subsec:search-routing} resolves the majority of real queries with one or two navigation steps.

\paragraph*{Answer quality.}
The mean human rating of 2.86---between \emph{related} (2) and \emph{exact hit} (3), and much closer to the latter---shows that on real author-knowledge-base queries \textsc{WikiKV} most often returns directly usable answers rather than merely related material. Together with the latency profile above, this provides real-world evidence complementing the controlled \textsc{AuthTrace} evaluation.

\subsection{Scalability}
\label{subsec:scalability}
\begin{figure}[htbp]
\centering
\includegraphics[width=\linewidth]{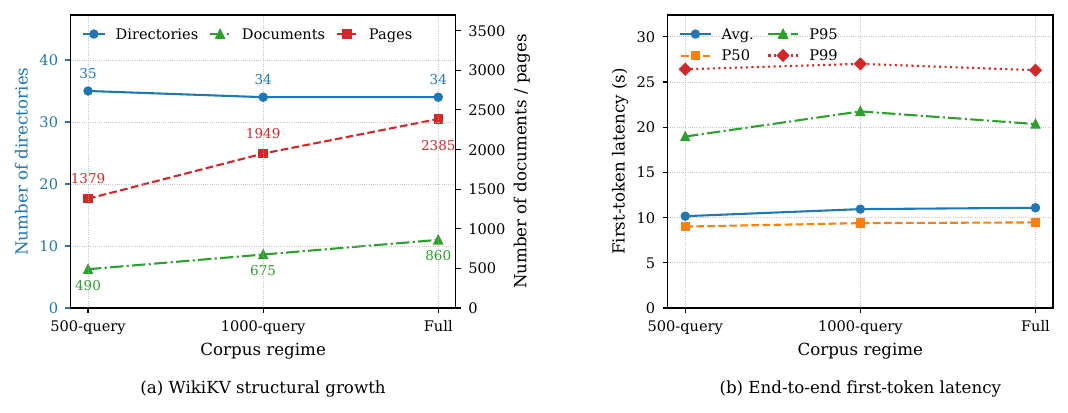}
\caption{End-to-end scalability of \textsc{WIKIKV} on \textsc{AuthTrace}.
}
\label{fig:scalability}
\end{figure}

We sample three nested regimes from \textsc{AuthTrace} of increasing corpus size: \textbf{500-query}, \textbf{1000-query}, and \textbf{full}. For each we record the structural footprint of the induced instance (directories and pages) and the first-token latency at Avg./P50/P95/P99. Figure~\ref{fig:scalability}(a) plots structural growth---directory count essentially invariant while pages grow nearly proportionally with corpus size; Figure~\ref{fig:scalability}(b) plots the latency profile---the body drifts only mildly while the tail compresses. The two views jointly separate the effect of \emph{schema depth} from that of \emph{page mass} on latency.

\paragraph*{Sub-linear scaling of end-to-end latency.}
The latency curves of Figure~\ref{fig:scalability}(b) make the sub-linear scaling visually explicit: doubling the question set lifts the page count by $\sim$1.4$\times$ yet inflates average end-to-end latency by only $12.2\%$ and P99 latency by only $10.2\%$, and moving on to the full corpus adds less than $1\%$ on top. The first-token Avg.\ curve drifts only mildly ($+9.2\%$ across the three regimes) while P50 stays essentially fixed; the high-percentile metrics tighten in the same proportion, with P95 even contracting on Full and P99 confined to a $0.7$\,s band. Scale therefore enters the system through the body of the latency distribution rather than through tail blow-ups.

\paragraph*{Why the growth is nearly flat.}
The shape follows directly from the model of \S\ref{sec:storage-model}--\S\ref{sec:navigation-query}. Every query's Q1 point lookup is $O(1)$ in the KV-pair count by the path-as-key co-location of \S\ref{sec:storage-model}, a constant at every scale; the subsequent path navigation is bounded by schema depth $D$, not by $|V|$, and the cold-start of \S\ref{subsec:schema-cold-start} keeps $D$ stable as the corpus grows. The flat ``Directories'' curve in Figure~\ref{fig:scalability}(a) confirms this---the directory count holds at $34$--$35$ while documents and pages grow $\sim$1.75$\times$ and $\sim$1.73$\times$. The residual $\sim$10\% growth is thus per-page reading at the leaves and slightly larger \textsc{NeedsDeeper} candidate sets, not navigation depth.

\subsection{Ablation}
\label{subsec:ablation}

To isolate the two most distinctive design choices in \textsc{WikiKV}, namely the sampled cold-start of \S\ref{subsec:schema-cold-start} and the search-routing operator of \S\ref{subsec:search-routing}, we run an end-to-end ablation on the single-author \textsc{Lu Xun} corpus, the densest thematic subset of \textsc{AuthTrace}. We compare three configurations under an identical query workload, generation model, and prompt template: \textbf{Full} (the complete system); \textbf{w/o Cold-Start}, which injects the full document set into schema construction instead of the sampled subset; and \textbf{w/o Search Routing}, which disables the Phase-1 router and falls back to pure layer-by-layer navigation.

\begin{table}[htbp]
\caption{Ablation on the \textsc{Lu Xun} Corpus (the densest thematic subset of AUTHTRACE): Removing Cold-Start Sampling or Search Routing}
\label{tab:ablation-luxun}
\centering
\renewcommand{\arraystretch}{1.15}
\resizebox{\linewidth}{!}{%
\begin{tabular}{l|c|c|c}
\toprule
\textbf{Metric} & \textbf{Full} & \textbf{w/o Cold-Start} & \textbf{w/o Search Routing} \\
\midrule
Avg.\ tool calls / query      & 3.53 & 3.75 & 5.42 \\
Avg.\ pages read / query      & 1.56 & 1.63 & 5.40 \\
\midrule
Avg.\ first-token latency (s) & \textbf{11.9} & 14.8 & 15.1 \\

AC ($\uparrow$)               & \textbf{65.6} & 54.7 & 40.2 \\
\bottomrule
\end{tabular}
}

\end{table}
\paragraph*{Cold-start sampling vs.\ full-document injection.}
Replacing sampled cold-start with full-document injection (\textsc{w/o Cold-Start}) costs \textbf{10.9 absolute points} of AC (54.7 vs.\ 65.6) and degrades every latency metric: average first-token latency rises $+24\%$ (11.9\,s$\to$14.8\,s)(Table~\ref{tab:ablation-luxun}). The mechanism matches \S\ref{subsec:schema-cold-start}: feeding every document into schema induction inflates the prompt, biases the LLM toward over-specific topics from incidental overlaps, and yields a noisier, less discriminating schema. Downstream, tool calls (3.75 vs.\ 3.53) and pages read (1.63 vs.\ 1.56) rise as the navigator compensates online for a schema that was over-fit at construction time. Sampled cold-start thus delivers a strictly better schema for both accuracy and latency, not merely a cheaper one.

\paragraph*{Search routing vs.\ pure layer-by-layer navigation.}
The average first-token latency of \textbf{11.9\,s} shows the system delivers correct answers efficiently. Removing search routing causes \textbf{more than triple the pages read per query} (5.40 vs. 1.56) and \textbf{53\% more tool calls} (5.42 vs. 3.53), reflecting the deeper navigation required when the router cannot prune irrelevant branches. Latency increases modestly to 15.1 s, which is a side effect of the extra exploration rather than a primary cost. The real penalty is accuracy: without search routing the AC drops by \textbf{25.4 points} (65.6\% → 40.2\%). The system thus allocates navigation budget precisely where it matters—fewer pages, fewer tool calls, and preserved correctness for the same latency order of magnitude. This validates the search-accelerated design: routing quality, not raw speed, is the bottleneck.

\section{Related Work}
\label{sec:related-work}
We review existing work on hierarchical data management, graph and key--value databases for knowledge, and retrieval-augmented generation.

\subsection{Hierarchical Data Management}
\label{subsec:hierarchical-data}

Directory services such as LDAP~\cite{rfc4511ldap} organize entries under a globally unique DN hierarchy. Distributed file systems such as GFS~\cite{ghemawat2003gfs} and HDFS~\cite{shvachko2010hdfs} provide a parent--child layout but expose only opaque byte streams at the leaves and treat directory metadata as bulk-oriented bookkeeping rather than a payload-bounded, queryable listing primitive. Warehouse-style query layers such as Hive~\cite{thusoo2009hive} sit above these file systems, but their relational, set-oriented evaluation model is again ill-matched to the page-and-enumerate-children contract of LLM-driven navigation.

More closely related is native and relational treatment of XML. TIMBER~\cite{jagadish2002timber} pioneered set-at-a-time evaluation of tree queries via structural joins, and Tatarinov~\cite{tatarinov2002xml} encoded ordered XML into relations without losing XPath expressiveness. However, all of the above systems expose navigation interfaces that return entire subtrees or unbounded child sets, with no mechanism to cap the payload size of a single traversal step under an LLM's context window.

In contrast, \textsc{WikiKV} provides budget-aware hierarchical navigation tailored to LLM-driven traversal (\S\ref{sec:navigation-query}).

\subsection{Graph Databases and Key--Value Stores for Knowledge}
\label{subsec:graphdb-vs-kv}

Property graph databases like Neo4j with Cypher~\cite{angles2017graph,francis2018cypher,robinson2018graphdatabases} excel at multi-hop reasoning over heterogeneous edges and power large knowledge graphs such as Freebase~\cite{bollacker2008freebase}, DBpedia~\cite{auer2007dbpedia}, and Wikidata~\cite{vrandecic2014wikidata}. Yet none exposes a native ``directory listing'' primitive: enumerating children needs Cypher patterns or labelled traversals whose planner overhead is disproportionate to an $O(k)$ prefix scan.

At the opposite end, industrial KV and wide-column stores such as Dynamo~\cite{decandia2007dynamo}, Bigtable~\cite{chang2008bigtable}, Cassandra~\cite{lakshman2010cassandra}, and Spanner~\cite{corbett2013spanner}---typically atop LSM-tree backends~\cite{oneil1996lsm}---provide $O(1)$ lookups and scalability but are deliberately \emph{flat}: no hierarchical key space, no contract tying a parent's value to its children's existence. Building a wiki on them forces the application to reinvent path semantics, the gap \textsc{WikiKV} closes by lifting the hierarchy into the key encoding. \textsc{WikiKV} is thus not a replacement for graph or KV stores but a thin \emph{path-indexing layer} over any backend supporting point lookups and prefix scans (\S\ref{sec:storage-model}).

\subsection{Retrieval-Augmented Generation and Knowledge Retrieval}
\label{subsec:rag-retrieval}

The dominant paradigm for grounding LLMs~\cite{openai2023gpt4,yang2024qwen2,grattafiori2024llama3} in external corpora is RAG~\cite{lewis2020rag,gao2023ragsurvey}: a dense retriever like DPR~\cite{karpukhin2020dpr}, ColBERT~\cite{khattab2020colbert}, or REALM~\cite{guu2020realm} over an ANN index such as HNSW~\cite{malkov2020hnsw}, FAISS~\cite{johnson2021faiss}, or Milvus~\cite{wang2021milvus}, with readers like Fusion-in-Decoder~\cite{izacard2021fid} aggregating evidence. Unlike these methods, which flatten the corpus into a bag of chunks and force the LLM to reconstruct discarded structure, \textsc{WikiKV} answers by walking the hierarchy rather than concatenating top-$k$ chunks.

Recent work recovers structure inside the retriever: RAPTOR~\cite{sarthi2024raptor} clusters and summarizes chunks into a balanced tree; MemWalker~\cite{chen2023memwalker} treats long-context reading as navigation over a summary tree; GraphRAG~\cite{edge2024graphrag} extracts an entity-relation graph with community summaries; and agentic frameworks like ReAct~\cite{yao2023react} interleave retrieval with reasoning. 

These operate over a static in-memory hierarchy and leave schema evolution unaddressed; \textsc{WikiKV} adds storage-layer evolution operators (\S\ref{sec:schema-evolution}) that let the hierarchy evolve safely under concurrent reads and writes.

\section{Conclusion}
\label{sec:conclusion}


In this paper, we present \textsc{WikiKV}, a path-indexed key--value
storage model purpose-built for LLM-curated hierarchical knowledge
bases, deployed at scale on the WeChat Official Account platform.
By materializing the wiki schema directly into the key namespace,
every directory listing reduces to a single point lookup in $O(1)$
storage round-trips rather than a depth-dependent traversal. On top
of this, a data-driven schema layer cold-starts via Intent-Anchored
Schema Induction and continuously evolves through mutual-information
merge and Architect--Critic--Arbiter split, while a parent-after-child
write protocol rules out partial reads under concurrent offline
rewrites without locking the read path, and a budgeted navigation
operator compresses LLM-assisted descent from $O(\mathit{depth})$ to
$O(1)$. Experiments on \textsc{AuthTrace} and online deployment in the
WeChat Official Account AI Assistant confirm consistently low per-operator
latency across relational,
graph, and FS backends, and superior end-to-end answer correctness relative to RAG baselines. Future work targets multimodal knowledge bases, adaptive depth tuning, stronger consistency under concurrent offline writers, and cross-domain generalization beyond AUTHTRACE's single literature domain.

\section*{AI-Generated Content Acknowledgement}

We disclose the use of artificial intelligence (AI) in the experimental evaluation reported in Section~\ref{sec:experiments} as a drop-in inference component. Specifically, \textsc{DeepSeek-V4-Flash} was used in three places: (i) the offline construction of the \textsc{WikiKV} instance with Intent-Anchored cold-start schema induction; (ii) the online schema-evolution pipeline; and (iii) the end-to-end retrieval-augmented question-answering pipeline of \textsc{WikiKV} and all RAG baselines. 


\bibliographystyle{IEEEtran}
\bibliography{references}

\end{document}